\documentclass[prl,twocolumn,superscriptaddress,showpacs,preprintnumbers,amsmath,amssymb]{revtex4}
\usepackage{amsmath}
\usepackage{graphicx}
\usepackage{dcolumn}

\begin{document}

\title{Dielectric mismatch and shallow donor impurities in GaN/HfO$_2$ quantum wells}

\author{T. A. S. Pereira}
\affiliation{Institute of Physics, Federal University of Mato Grosso, 78060-900, Cuiab\'a, Mato Grosso, Brazil}
\author{A. A. Sousa}
\affiliation{Department of Physics, Federal University of Cear\'a, C.P. 6030, 60455-900, Fortaleza, Cear\'a, Brazil}
\author{M. H. Degani} 
\affiliation{Universidade Estadual de Campinas, 13484-350, Limeira, S\~ao Paulo, Brasil}
\author{G. A. Farias}
\affiliation{Department of Physics, Federal University of Cear\'a, C.P. 6030, 60455-900, Fortaleza, Cear\'a, Brazil}

\begin{abstract}
In this work we investigate electron-impurity binding energy in GaN/HfO$_2$ quantum wells. The calculation considers simultaneously all energy contributions caused by the dielectric mismatch: (i) image self-energy (i.e., interaction between electron and its image charge), (ii) the direct Coulomb interaction between the electron-impurity and (iii) the interactions among electron and impurity image charges. The theoretical model account for the solution of the time-dependent Schr\"odinger equation and the results shows how the magnitude of the electron-impurity binding energy depends on the position of impurity in the well-barrier system. The role of the large dielectric constant in the barrier region is exposed with the comparison of the results for GaN/HfO$_2$ with those of a more typical GaN/AlN system, for two different confinement regimes: narrow and wide quantum wells. 
\end{abstract}

\maketitle

\section{Introduction}
\label{intro}

When an impurity is introduced into a low dimensional structure, such as quantum wells (QW), nano wires (NW) and quantum dots (QD) the calculation of the electronic properties in this structures becomes considerably more complex if compared to that of a doped three-dimensional crystal \cite{bin,jarosik,weber,fraizzoli,ferreyra,fpeeters}. This occurs because of the restricted movement in the structure growth direction, which is imposed by the potential due to band edges discontinuities $\Delta E$. First, the binding energy of the carrier-impurity in the structure depends on the confinement potential $\Delta E$, and second, both the binding energy and wave function of the carrier and impurity depends on the impurity position in the structure growth direction. On the other hand, due to recent progress in epitaxial crystal growth techniques, such as molecular beam epitaxy (MBE), research focusing on impurity and electronic states in nano-structures has attracted great attention \cite{morgan,vinh}. However, effects caused by image charges due to the dielectric mismatch at the structure interface have been overlooked. Indeed, donor binding energy can be significantly modified by additional confinement effects that image charges distribution produce  \cite{cen,deng}. Thus, recent research  focusing on high-k dielectrics based QWs and NWs reveals interesting results related to carrier confinement \cite{teldo,teldo1, teldo2, teldo3}. We recently demonstrate that the interaction between carriers and their image charge, induced by the dielectric mismatch ($\varepsilon _r = \varepsilon _{\rm{GaN}}/ \varepsilon_{\rm{HfO}_2}$ = 9.5 / 25 = 0.38), strongly modifies the electronic structure in GaN/HfO$_2$ QWs (and NWs)  and for wide QWs (wide radii NWs) heavy holes are confined in interfacial regions, similar to that observed in type-II heterostructures \cite{teldo2, teldo3}.  Such interfacial confinement leads to drastic modiﬁcations on the electronic properties of the QWs and NWs. Particularly, for NWs under an applied magnetic field, where angular momentum transitions occur in the ground state due to the Aharonov-Bohr effect \cite{teldo3}. A decrease in the oscillator strength of electron-hole pairs in $\varepsilon _r < 1$ QWs and NWs is also predicted for wide QW and larger wire radii, which directly affects their recombination rates \cite{teldo2, teldo3}.

In this work, we investigate electron-impurity binding energy in GaN/HfO$_2$ Qws. As for illustration we compared this results with those of a more typical AlN/GaN system. The presence of a point charge in a region where the dielectric constant is discontinuous induces polarization charges at the QW interfaces, and this problem can be solved by the image charges method \cite{stern}. As shown here, The electron energy, electron wave function and the electron-impurity binding energy can change significantly due to additional confinement effects produced by the image charge distribution. Our calculation considers simultaneously all energy contributions caused by the dielectric mismatch: (i) image self-energy (interaction between electron and its image charges), (ii) the direct Coulomb interaction between electron and the actual impurity, as well as (iii) the interactions among electron and impurity image charges. Moreover, from practical means, we also investigate stark effect and electron-impurity binding energy for two different confinement regimes: narrow (5 nm) and wide (10 nm) QWs. When compared to the effective Bohr's radius for the GaN bulk $a_B^{\star}$ ($ = a_B \varepsilon_{\rm{GaN}} / m_e^\ast = 2.65$ nm; where $a_B = 0.53 \AA $ is the Bohr's radius) narrow and wide QWs used in this work are twice and four times the effective Bohr's radius, respectively. The binding energy of an electron bound to a hydrogenic impurity is obtained as function of the impurity position, by solving a fully three-dimensional time dependent Schr\"odinger equation using a method with neither adjustable parameters nor restrictive basis expansions as employed by almost all theoretical approaches in the literature \cite{Degani0,degani1,degani2}. For simplicity, we address zinc blende GaN instead of its wurtzite crystalline structure in order to avoid more complicated polarizations effects observed in this phase \cite{vurgaftman}.

\section{Theoretical Model}
\label{sec:1}

\textit{1. Time-dependent Schr\"odinger equation}\\

The theoretical method used to calculate the binding energy of an electron bound to a hydrogenic impurity is based on the adiabatic approximation. The time-dependent Schr\"odinger equation \cite{kosloff, degani, leburton, Chaves} is consistent with the effective mass approach and the envelope function formalism  

\begin{equation}
\label{eq1} i\hbar \frac{\partial }{{\partial t}}\psi \left( {r,t} \right) = H\psi \left( {r,t} \right),
\end{equation}
\vspace{0.01cm}

\noindent and describes the quantized states of a single particle coupled to a quantum well under the effect of impurity Coulomb potential and potential due to image charges. The Hamiltonian $H$ is given by 

\begin{equation}
\label{eq2}H=\frac{1}{2}P\frac{1}{m^*(r)}P
+ V(r),
\end{equation}
\vspace{0.01cm}

\noindent where $P = -i\hbar \nabla$ is the kinetic energy operator and $V(r)$ is the potential energy operator. The initial solution $\Psi(r,t)$ given by the method is

\begin{equation}
\label{eq3}\Psi (r,t)= \exp{ \left( { - \frac{i}{\hbar }\int\limits_0^t {Hdt} } \right)}\Psi (r,0).
\end{equation}
\vspace{0.01cm}

\noindent The Hamiltonian of Equation ~(\ref{eq2}) does not depend on time, so the integral in Equation ~(\ref{eq3}), solved in the range between $t$ and $t + \Delta t$ is given by

\begin{equation}
\label{eq4}\Psi(r,t+\Delta t)=\exp{ \left({- \frac{i}{\hbar }H\Delta t} \right)}\Psi (r,t),
\end{equation}
\vspace{0.01cm}

\begin{figure}[t]
\begin{center}
\centering
\includegraphics[scale=0.68]{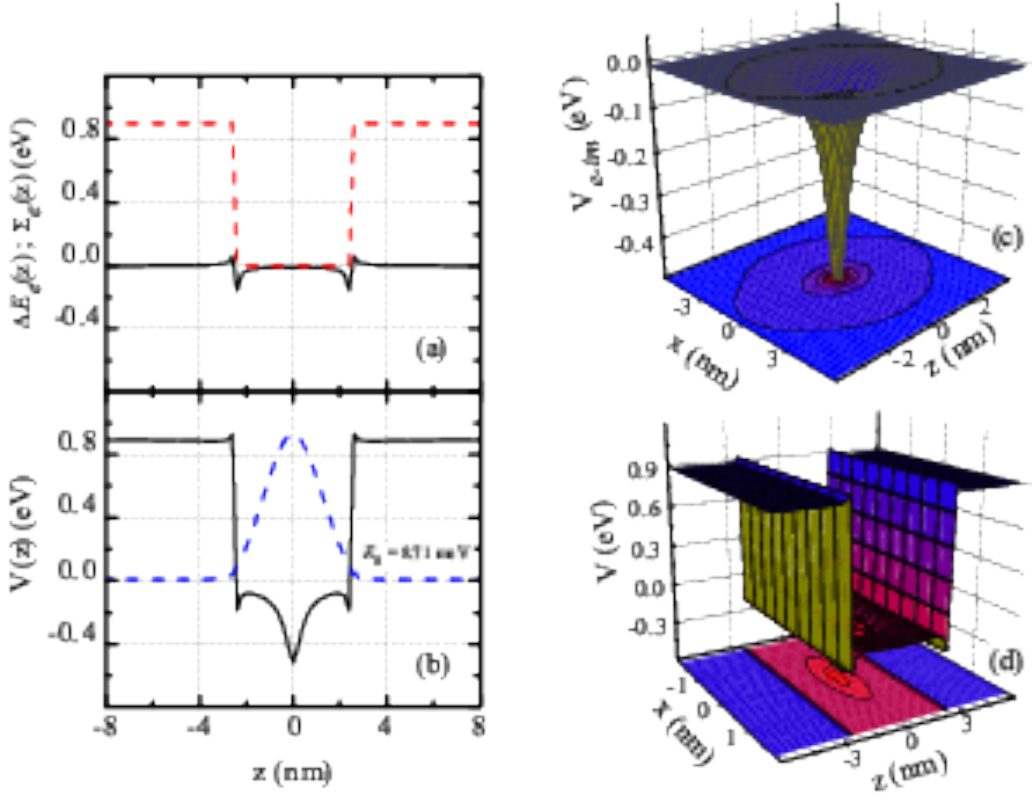}
\caption{\small{\label{fig1} (a) Energy potential $\Delta E_e(z_e)$ due to conduction band edge descontinuity (red dashed line) and the potential $\Sigma_e(z_e)$ due to self-energy corrections (black solid line). (b) Total potential $V\left( r \right) = \Delta {E_e}\left( {{z_e}} \right) + {\Sigma _e}\left( {{z_e}} \right) + {V_{e - im}}\left( r \right)$ in the $z$ direction (black solid line) and electron ground state wave function (blue dashed line). (c) Coulomb potential $V_{e-m} (r)$ of electron-impurity interaction in 3D plot. (d) Total potential $V(r)$ in 3D plot.}}
\end{center}
\end{figure}

\noindent which is approximated by the expression

\[
\Psi \left( {r,t + \Delta t} \right) = {\exp{  \left[- {iV\left(r \right)\Delta t/2\hbar } \right]}}
\]

\[\times {\exp{  \left[ -{i{p^2}\Delta t/2\hbar m^*} \right]}} \]
\begin{equation}
\label{eq5} \times {\exp{ \left[- {iV\left( r \right)\Delta t/2\hbar } \right]}}+O \left( \Delta t^3 \right).
\end{equation}
\vspace{0.01cm}

The error introduced in this expression, when we drop the term $O \left( \Delta t^3 \right)$, results from the noncommutability of the kinetic and potential operators. The potential operator $V\left( r \right)$, with $r = \left( {{\rho _e},{z_e},{z_{im}}} \right)$ and $\rho=\sqrt{x^2+y^2}$, is given by

\begin{equation}
\label{eq6}V\left( r \right) = \Delta {E_e}\left( {{z_e}} \right) + {\Sigma _e}\left( {{z_e}} \right) + {V_{e - im}}\left( r \right),
\end{equation}
\vspace{0.01cm}

\noindent where $\Delta E_e\left( z_e \right)$ is the heterostructure band edge confinement, $\Sigma_e\left( z_e \right)$ is the self-energy potential and $V_{e - im} \left(r \right)$ is the direct electron-impurity Coulomb interaction. The last term includes direct eletron-impurity Coulomb interaction and the interactions between electron and impurity image charges. This contribution to the total potential was deduced from solutions of the Poisson equation in 2D quantum structures, as shown in Eqs. (A21)-(A25) of Reference \cite{kumagai}. Fig. ~\ref{fig1} shows each potential given by Equation (\ref{eq6}), for a 5 nm QW: Fig. ~\ref{fig1}(a) shows the potential due to band edges confinement $\Delta E_e\left(z_e \right)$ (red dashed line) and the self-energy potential $\Sigma_e(z_e)$ (black solid line), which is attractive (repulsive) for charge on the low (high) dielectric constant side ($\varepsilon_{\rm{GaN}} < \varepsilon_{\rm{HfO}_2}$). The attractive potential on the well region produce cusps that appears near the edges of the interface transition layers, shown in the total potential depicted in Fig. \ref{fig1}(b).  For the pourpose of our analyses, we plot in Fig. ~\ref{fig1}(c) and ~\ref{fig1}(d) the potential $V_{e - im} \left( r \right)$, due to direct electron-impurity Coulomb interaction, and the total potential $V\left(r \right)$ in a three-dimensional space, respectively.

The eigenstates of the Hamiltonian are calculated by usin a propagation scheme in the imaginary time domain. \cite{leburton} Any wave function can be written as a linear combination of the eigenstates of a Hamiltonian, since it forms a complete orthogonal basis

\begin{equation}
\label{6.1}{\left| \Psi  \right\rangle _t} = \sum\limits_{n = 0}^\infty  {{a_n}{e^{ - \frac{{i{E_n}t}}{\hbar }}}\left| {{\varphi _n}} \right\rangle } ,
\end{equation}

\noindent where $\varphi _n$ and $E_n$ are the eigenfunction and eigenenergy of the $nth$ eigenstate, respectively. Using $\tau=it$,

\[
{\left| \Psi  \right\rangle _t} = \sum\limits_{n = 0}^\infty  {{a_n}{e^{ - \frac{{{E_n}\tau }}{\hbar }}}\left| {{\varphi _n}} \right\rangle } 
\]
\begin{equation}
\label{6.2} = {e^{ - \frac{{{E_0}\tau }}{\hbar }}}\left[ {{a_0}\left| {{\varphi _0}} \right\rangle  + \sum\limits_{n = 1}^\infty  {{a_n}{e^{ - \frac{{\left( {{E_n} - {E_0}} \right)\tau }}{\hbar }}}\left| {{\varphi _n}} \right\rangle } } \right].
\end{equation}

\noindent After several imaginary-time steps of propagation ($\tau \rightarrow \infty$), the term of the ground state, ${e^{ - \frac{{{E_0}\tau }}{\hbar }}}{a_0}\left| {{\varphi _0}} \right\rangle$, becomes strongly dominant over the terms of the sum, since $E_n-E_0> 0$ for $n> 0$. Therefore, starting with any wave function, this function should converge to the ground state of the system as $\tau$ increases. We can consider as a very long time those in which $\tau \gg \hbar / \left( E_n - E_0 \right)$. The excited states are obtained adding to the algorithm the Gram-Schmidt orthonormalization method which will assure orthonormality between all states in each time step.
\\

\textit{2. Self-energy potential}\\

In order to calculate the effects of the self-energy potential $\Sigma_e(z_e)$, shown in Fig. ~\ref{fig1}(a) (black solid line), on the electron energy we use the method based on image charges. The electrostatic potential due to a charge $Q$ located at $r = (0,0,z_0)$, in a medium where the dielectric constant $\varepsilon (z)$ depends on the position is given by

\begin{equation}
\label{eq7}\nabla  \cdot \left[ {\varepsilon (z)\nabla \phi (r)} \right] =  - Q\delta (r - r_0 ).
\end{equation}
\vspace{0.01cm}

The solution in cylindrical coordinates is independent of the azimuth angle (see detail in References \cite{stern,teldo2}). In this case, we can write $\phi(r)$ in the general series as

\begin{equation}
\label{eq8}\phi ( r ) = \int\limits_0^\infty  {qJ_0 (qR)A_q (z)dq},
\end{equation}
\vspace{0.01cm}

\noindent where $J_0 (qR)$ is the Bessel function of the zeroth order, $A_q (z)$ is a function determined by the boundary conditions of $\phi (r)$ at the interfaces. The solution for the image self-energy potential $\Sigma _e (z_e )$ is

\begin{equation}
\label{eq9}\Sigma _e (z_e )= \frac{Q} {2}\int\limits_0^\infty  {q\left[ {A_q (z_0 ) - A_q^0 (z_0 )} \right]dq},
\end{equation}
\vspace{0.01cm}

\noindent where $A_q^0 (z_0 )$  is solution of Equation (\ref{eq8}) if $\varepsilon $ is $z$ independent. Without loss of generality, we shall here consider QWs with abrupt interface. The self energy potential $ \Sigma_e(z_e)$ diverges at the interface $z= \pm L/2$ and we employ a numerical grid such that the coordinate in $z= \pm L/2$ does not sit at a grid point in order to avoid the divergence problem. The major results for $\Sigma_e(z_e)$ can be seen in the Reference \citep{teldo2} and will not be repeat here.

\section{Results and Discussion}

\begin{figure}[t]
\begin{center}
\centering
\includegraphics[scale=0.7]{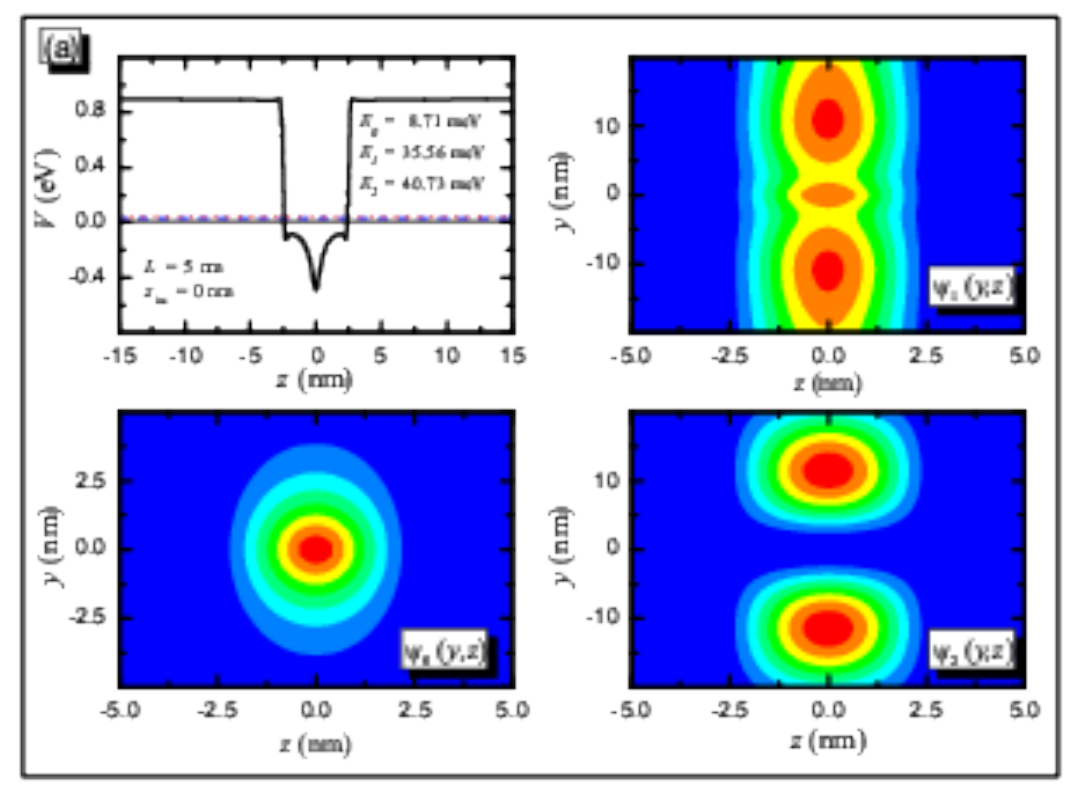}
\includegraphics[scale=0.7]{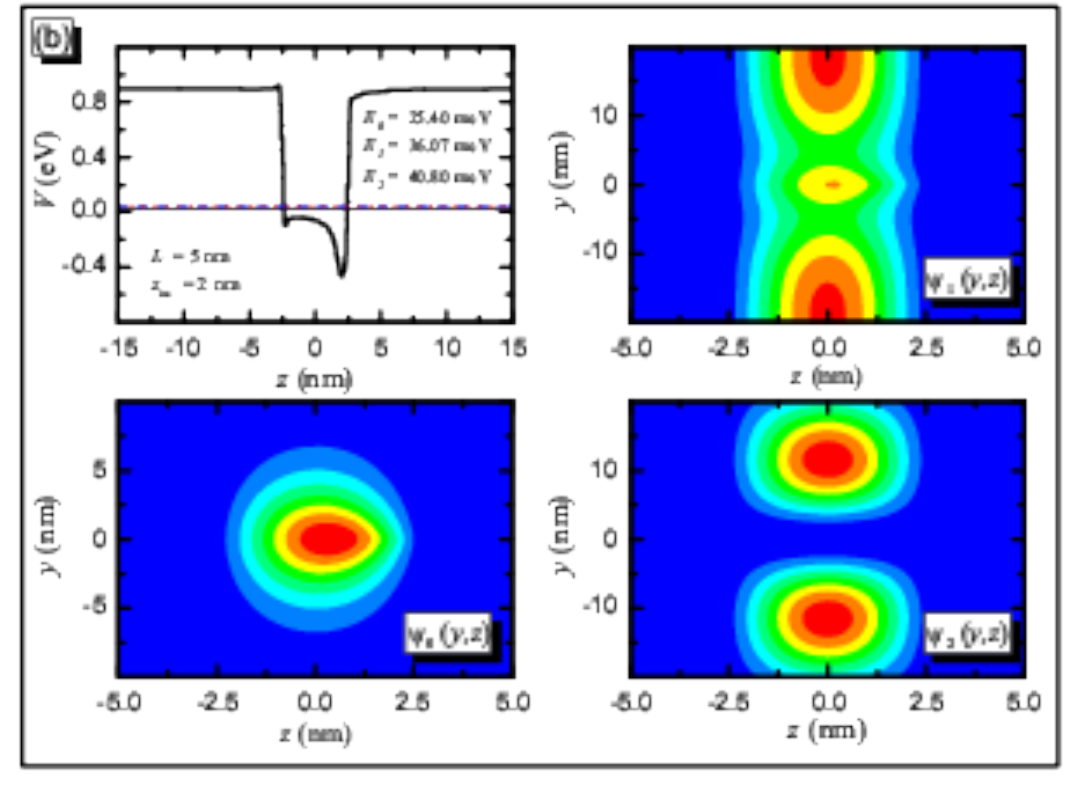}
\caption{\small{\label{fig2}(Color online) Waves functions projection in the $(yz)$ plane for the ground state, first and second excited states. In (a) the impurity is located in $z_{im}= 0$ nm and in (b) the impurity is $z_{im}= 5$ nm far from the center of the QW. The z-projection of the total potential $V(z)$, in eV, are depicted for QWs with width of $L = 5$ nm. }}
\end{center}
\end{figure}

As in the case of a model structure, QWs are formed by a zinc blende GaN \thinspace layer ranging in the region $|z|\leq a$ between two \thinspace HfO$_2$ layers in the region $|z|\geq a$. Between these materials, we consider the existence of abrupt interfaces at $a$ position along the $z$ axis. The GaN electron effective mass were taken from experiments ($m_e^{\ast} = 0.19$) \cite{vurgaftman}, and for simplicity, we have considered the electron effective mass invariable along $z$. Although photoemission spectroscopy experiments demonstrated that $\Delta E_e=2.1$ eV for wurtzite GaN/HfO$_2$  interfaces \cite{cook}, the absence of this information for the zinc blende heterojunction leads us to estimate these quantities through the simple electron affinity model \cite{korotkov,wu1}, for which we obtain $\Delta E_e = 0.9$ eV. As shown in Fig. ~\ref{fig1}, the quantum well has mirror symmetry from the origin in the $z$ direction, at $z = 0$, and the reference of the total potential $V (z)$ in Equation ~(\ref{eq6}) is taken with respect to the zero level of the potential $\Delta E_e (z)$, as shown in Fig. ~\ref{fig1}(a).

\begin{figure}[t]
\begin{center}
\centering
\includegraphics[scale=0.71]{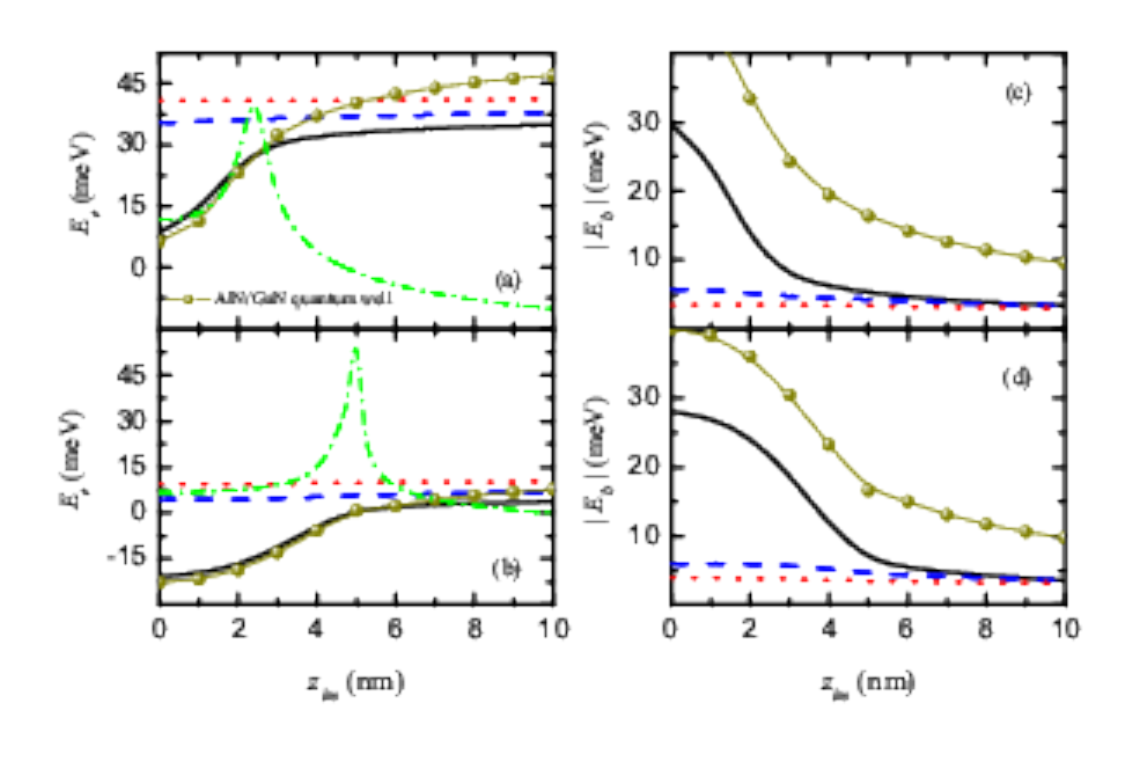}
\caption{\small{\label{fig3}(Color online) Left panels: Electron energy for ground state (black solid line), first (blue dashed line) and second (red doted line) excited state in QW for (a) narrow $L=5$ nm QW width and (b) wide $L=10$ nm HfO$_2$/GaN QW width. Right panels: Electron-impurity binding energy for ground state (black solid line), first (blue dashed line) and second (red dotted line) excited state energy as function of impurity position for a (c) narrow ($L=5$ nm) QW and (d) wide $L=10$ nm HfO$_2$/GaN QW. The dark yellow line-sphere depict the electron energy (left) and electron-impurity binding energy (right) in narrow (top) and wide (bottom) AlN/GaN QW, and the green dash-dot line shows the effect of the image charges in GaN/HfO$_2$ QW.}}
\end{center}
\end{figure}

The impurity can be placed at any position along $z$ direction, and two particular positions are show in the Fig. \ref{fig2}. Fig. ~\ref{fig2}(a) depicted the total potential $V(z)$ projeted along $z$ direction, where the impiruty is located at the center of the QW in $(x_{im},y_{im},z_{im})=(0.0,0.0,0.0)$ nm. Fig. ~\ref{fig2}(b) shows the total potential $V(z)$ with the impiruty located at the interface in $(x_{im},y_{im},z_{im})=(0.0,0.0,2.5)$ nm.  These figures also display the energy and the projection of the electron wave function in the $(y,z)$ plane, for ground state $\psi_0(y,z)$, first $\psi_1(y,z)$ and second $\psi_2(y,z)$ excited states, confined in a 5 nm QW. For example, when the impurity is located in the center of the QW the ground state energy is about 8.71 meV upward of potential energy reference, and goes up to 25.40 meV when the impurity is placed at the interface of the QW. Noteworthy the potential energy $V(z)$ is attractive in the well region due to both electron-impurity interaction and attractive behavior of the image self-energy. This potential move the electron to the center of the QW and the wave function is concentrated in that region, as depicted in Fig. ~\ref{fig2}(a). When the impurity is located at the interface, for a 5 nm QW, the electron is pushed towards to the right interface, as shown by the ground state  $\psi_0(y,z)$ and first excited state $\psi_1(y,z)$ wave fuction. Interestingly, the second excited state $\psi_2(y,z)$ is weakly attracted by the impurity.

\begin{figure}
\begin{center}
\centering
\includegraphics[scale=0.42]{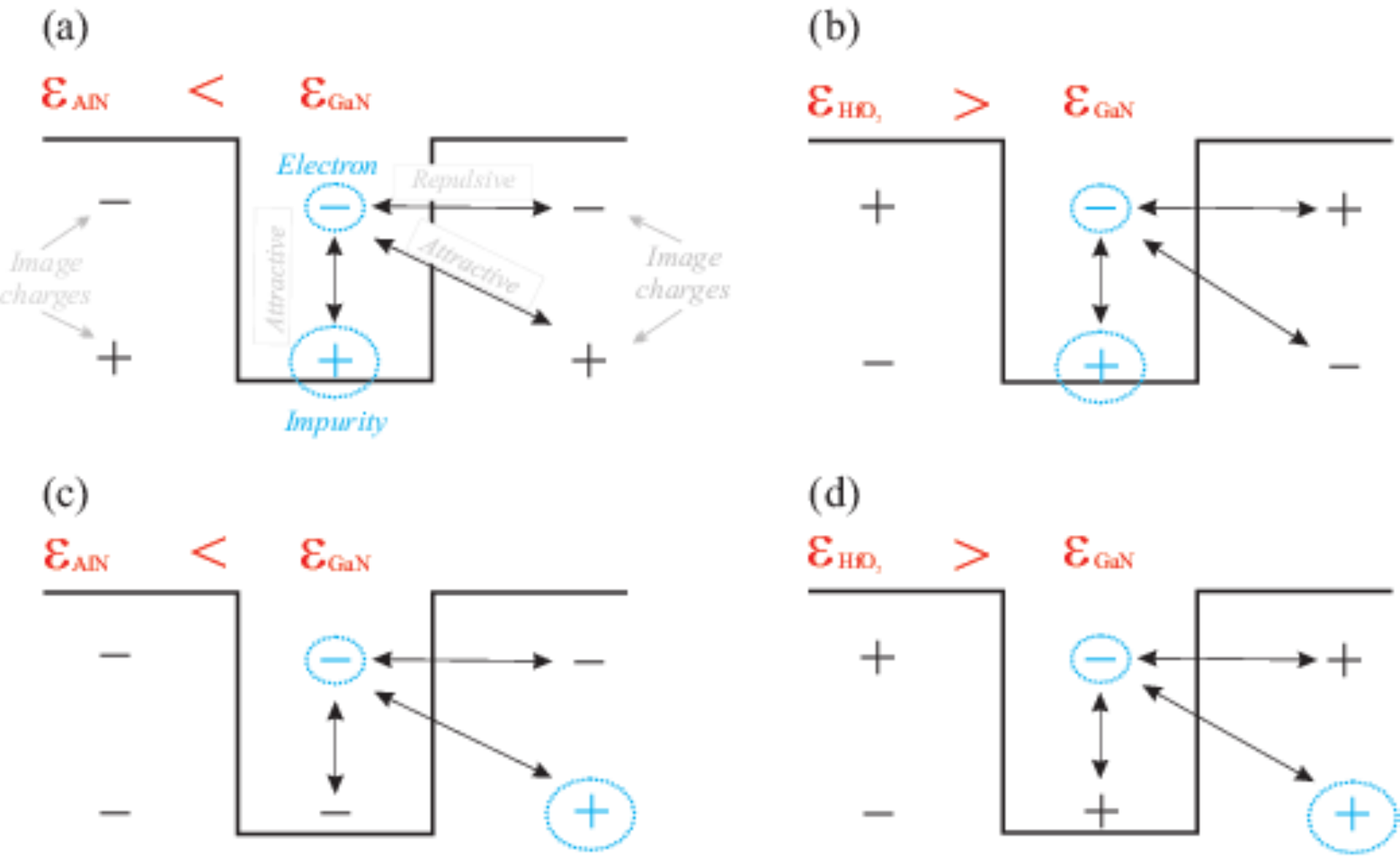}
 \caption{\small{\label{fig4} Schematic diagram of different interactions between electron and their image charges, electron and impurity as well as electron and impurity image charges for GaN/AlN (a)-(c) and in GaN/HfO$_2$ (b)-(d) QWs. In (a)-(b) the impurity is located in the well region while in (c)-(d) the impurity is located in the barrier region.}}
\end{center}
\end{figure}

Figures ~\ref{fig3}(a) and ~\ref{fig3}(b) ilustrate the electron energy as function of the impurity position along $z$ axis, in narrow ($L = 5$ nm) and wide ($L = 10$ nm) QWs, respectively, for the ground state energy (solid lines), first (dashed lines) and second (dotted lines) excited states. Our result shows that the ground state energy increases asymptotically until the point where it reaches values with less pronounced variations from $z_{im} \approx 2$ nm in narrow QW and from $z_{im} \approx 5$ nm in wide QW. For $z_{im}>2$ nm in narrow QW and $z_{im} > 5$ nm in wide QW the ground state energy is invariant with $z_{im}$ position, which indicates that the effect of the impurity potential is small when the impurity is located in the region of the barrier. Excited states are, on the other hand, weakly affected by the impurity position.

The \textit{n-th} electron-impurity binding energy level is calculated, with appropriate image charge contribution take into consideration, by the difference

\begin{equation}
\label{eq13}E_{n,b} = E_n(V_{e-im} \neq 0) - E_n(V_{e-im}  = 0),
\end{equation}
\vspace{0.01cm}

\noindent where the term $E_n(V_{e-im} \neq 0)$ means the $n$-th electron energy level calculated considering $V_{e-im}  \neq 0$ and $E_n(V_{e-im} = 0)$ is the $n$-th electron energy level calculated considering $V_{e-im}  = 0$, in Equation (\ref{eq4}). The absolute value of the electron-impurity binding energy, as function of the impurity position, is depicted in Fig. ~\ref{fig3}(c) and Fig. ~\ref{fig3}(d) for narrow ($L = 5$ nm) and wide ($L = 10$ nm) QWs, respectively. The curves are shown for the ground state (black solid line), first exited state (blue dashed line) and second exited state (red dotted line). As seen, the binding energy changes with impurity position in the QW structure. The maximum ground state electron-impurity binding energy value occurs in the center of the QW, for $z_{im} = 0$ nm and decreases when the impurity moves towards the interface (in both cases $L = 5$ nm and $L = 10$ nm) of the well region. For impurity at the interface region, the electron is weakly bound and the binding energy is about 5 meV, for the $z_{im}$ values investigated in this work. Excited states are always weakly bound to impurities, independent of the $z_{im}$ position.

\begin{figure}[t]
\begin{center}
\centering
\includegraphics[scale=0.8]{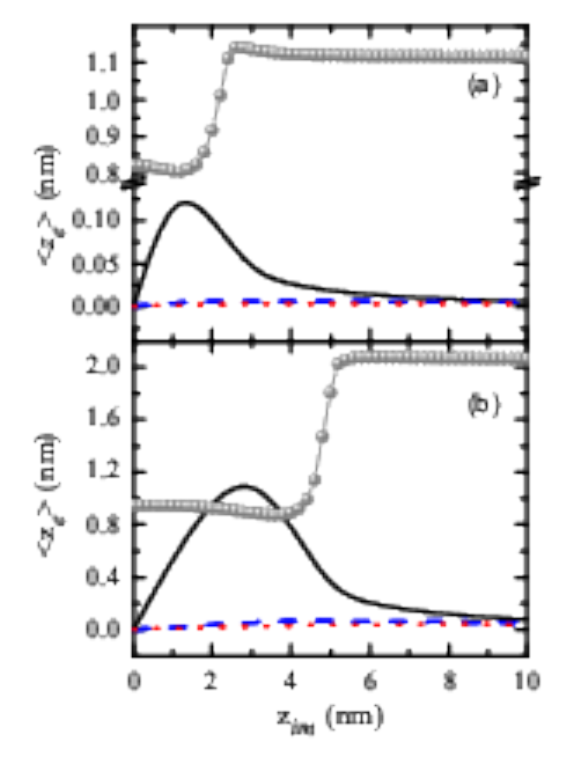}
 \caption{\small{\label{fig5} Electron center-of-mass as function of the impurity position ($z_{im}$) for (a) narrow ($L=5$ nm) and (b) wide ($L=10$ nm) QWs. The ground state, first and second excited states are represented by black solid, blue dashed and red dotted lines, respectively. The gray line-sphere depict the  standard deviation in position $\sigma_x$ for narrow and wide QWs.}}
\end{center}
\end{figure}

In order to help us to understand the role of the large dielectric constant in the barrier region we compare the results for GaN/HfO$_2$ with those of a more typical GaN/AlN system, where we have $\varepsilon _r = \varepsilon _{\rm{GaN}}/ \varepsilon_{\rm{AlN}}$ = 9.5 / 8.5 = 1.12. Figure \ref{fig3} show in dark yellow line-sphere the electron energy and electron-impurity binding energy in narrow and wide AlN/GaN QWs. Different from GaN/HfO$_2$ ($\varepsilon_r <1$) in a GaN/AlN ($\varepsilon_r > 1$) quantum well the electron fell a repulsive potential in the well region due to the dielectric mismatch. To elucidate the results presented in Figure \ref{fig3}(a) and (b) we show in Figure \ref{fig4} a schematic diagram of different interactions between electron, impurity and image charges for GaN/AlN (a)-(c) and GaN/HfO$_2$ (b)-(d) QWs. For the impurity at the well region, this picture clearly shows that the coulomb potential of impurity and image charges is more attractive in GaN/AlN QW compared to that in GaN/HfO$_2$ QW. For the impurity located at barrier region the coulomb potential becomes more attractive in GaN/HfO$_2$. This explains why the electron energy is smaller (larger) at the GaN/AlN system when the impurity is located in the well (barrier) region, as shown in Fig. ~\ref{fig3}(a) and (b). Without dielectric mismatch, or even for $\varepsilon_r > 1$, the confinement energy is always positive since the reference of confinement potential $V(z)$ is always either zero or larger. The energy $E_n(V_{e-im}  = 0)$ shown in the equation \ref{eq13} is bigger in GaN/AlN than that in GaN/HfO$_2$, giving rise to a difference in the binding energy as it is shown in Fig. ~\ref{fig3}(c) and (d). Aside from this difference this energy is essentially due to the band offset and the self energy potential, as it is shown in the Fig. ~\ref{fig1}(a). Further more, as the impurity position $z_{im}$ increase to the barrier region, the stationary states inside the well tends to discrete states analogous to the case of a quantum well without impurity, as we can see in Fig. ~\ref{fig3}(a) and (b) for $z_{im} >$ 5 nm. In the binding energy $E_{n,b}$ both contribution band offset and self energy potentials are not take into account and the states collapse near to $z_{im} =$ 10 nm, as shown in Fig. ~\ref{fig3}(c) and (d).

To clarify the role played by the high dielectric mismatch at the interfaces we add in the Fig. ~\ref{fig3} (a) and (b), in dash-dot green lines the difference in the electron energy take into account  image charges and does not take into account the image charges contributions for a GaN/HfO$_2$ quantum well. This results show essentially the contribution due to image self-energy (interaction between electron and its image charges) as well as the interactions among electron and impurity image charges.  As the impurity position increase this difference increase asymptotically until reach the maximum value around the interface position and decrease toward negative values in narrow quantum wells due to the attractive character of the self energy in systems with  $\varepsilon_r < 1$, as it can see on the Reference \citep{teldo1} and \citep{teldo2}.

To further elucidate here, the expectation value of the electron position $\left\langle {{z_e}} \right\rangle $, along of $z$ axis, is ploted as function of the impurity position $z_{im}$, for narrow ($L=5$ nm) and wide ($L=10$ nm) QWs in Fig. ~\ref{fig5}(a) and Fig. ~\ref{fig5}(b), respectively. For wells with $L = 5$ nm ($L = 10$ nm), the $\left\langle {{z_e}} \right\rangle $ of the ground state (solid lines) has maximum displacement around 0.12 nm (1.0 nm) when the impurity is located in $z_{im} = 1$ nm ($z_{im} = 3$ nm). Moving the impurity towards the barrier region, $\left\langle z_e \right\rangle $ tends to return to the QW center. In this case $\left\langle {{z_e}} \right\rangle$ of the excited states are also weakly affected by the impurity position. We also present in the Fig. \ref{fig5} in gray line-sphere the standard deviation in the position, namely the square root of the variance $\sigma _x = \sqrt {\left\langle z^2 \right\rangle  - {{\left\langle z \right\rangle }^2}}$. This quantity illustrate better the transition from strong binding to weak binding as $z_{in}$ goes into the barrier, illustrating the big variance at the interface position.

\begin{figure}
\begin{center}
\centering
\includegraphics[scale=0.9]{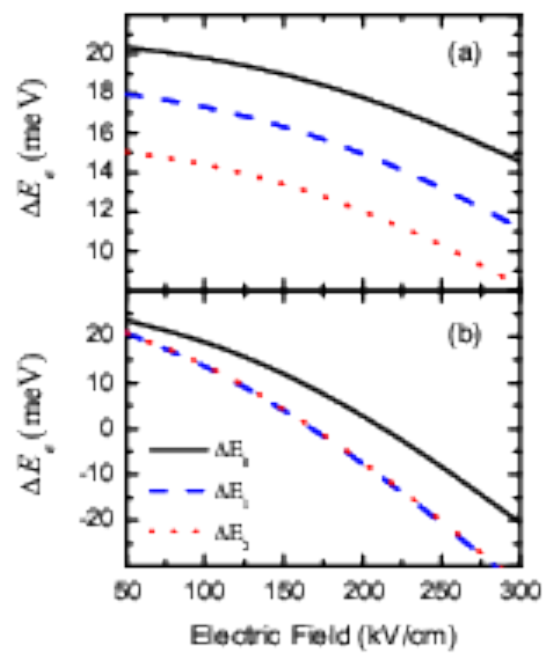}
 \caption{\small{\label{fig6} Stark shift of the ground state energy (solid lines), first excited state energy (dashed lines) and second excited state energy in (a) narrow (5 nm) and (b) wide (10 nm) QWs.}}
\end{center}
\end{figure}

Finally, from practical point of view, it is important to investigate the effects of external electric fields on the electronic structure of GaN/HfO$_2$ QWs. In Fig. \ref{fig6}, we show the stark shift $\Delta E_e = E_e(F\neq 0) -  E_e(F = 0)$ of the three first electron energy states, for (a) narrow and (b) wide QWs. The electron energy $E_e(F\neq 0)$ is calculated by considering an electric field $\mathbf{F}$, pointing along the $z$ direction, by including the term $eFz$ on Equation (\ref{eq6}). Here, it is important to notice that the shift on the electron energy $\Delta E_e$ can be underestimated by several meV with the applied electric field, i.e., $\simeq$ 6 meV for narrow QWs and $\simeq$  50 meV for wide QWs.

\section{Conclusions}

In conclusion, we have studied impurity state  with image charges effects in GaN/HfO$_2$ quantum wells. Our results show that the electron-impurity binding energy is highest when the impurity is located at the center of the quantum well and decreases when the impurity moves towards the interface. When the impurity is located on the barrier region the binding energy has a smaller  intensity. Moreover, when a electric field is applied on $z$ direction the electron energy can be shifted by about 50 meV in wide quantum wells. These results are lacking experimental confirmation, and we expect that our predictions induce the realization of such experiments.

\bigskip
\noindent
\textbf{Acknowledgments}
\bigskip T. A. S. Pereira was financially supported by PRONEX/CNPq/FAPEMAT 850109/2009 and by CAPES under process 3299/13-9.  A. A. Sousa has been financially supported by CAPES, under the PDSE contract BEX 7177/13-5. The authors also would like to acknowledge the CNPq, NanoBioEstruturas/CNPq, Pronex/FUNCAP/CNPq. Thanks to L. Craco for a careful reading of this manuscript.


\bigskip



\begin{thebibliography}{99}
\bibitem{bin}Bin Li, A. F. Slachmuylders, B. Partoens, W. Magnus, F. M. Peeters, Phys. Rev. B \textbf{77}, 115335 (2008).
\bibitem{jarosik}N. C. Jarosik, B. D. McCombe, B. V. Shanabrook, J. Comas, John Ralston and G. Wicks, Phys. Rev. Lett. \textbf{54}, 1283 (1985).
\bibitem{weber}G. Weber, Phys. Rev. B \textbf{41}, 10043 (1990).
\bibitem{fraizzoli} S. Fraizzoli, F. Bassani, Phys. Rev. B \textbf{41}, 5096 (1990).
\bibitem{ferreyra}J. M. Ferreyra and C. R. Proetto, Phys. Rev. B \textbf{52} 2309 (1995).
\bibitem{fpeeters}B. Li, B. Partoens, F. M. Peeters and W. Magnus, Phys. Rev. B \textbf{79} 085306 (2009).
\bibitem{morgan}C. G. Morgan, P. Kratzer, and M. Schefﬂer, Phys. Rev. Lett. \textbf{82} 4886 (1999).
\bibitem{vinh} N. Q. Vinh, H. Przybyli\'nska, Z. F. Krasil'nik, and T. Gregorkiewicz, Phys. Rev. Lett. \textbf{90}, 066401 (2003).
\bibitem{cen}  J. Cen and K. K. Bajaj, Phys. Rev. B \textbf{48}, 8061 (1993).
\bibitem{deng} Z. Y. Deng, and S. W. Gu, Phys. Rev. B \textbf{48}, 8083 (1993).
\bibitem{teldo}T. A. S. Pereira, J. A. K. Freire, V. N. Freire, G. A. Farias, L. M. R. Scolfaro, J. R. Leite, and E. F.  da Silva Jr., Appl. Phys. Lett. \textbf{88}, 242114 (2006).
\bibitem{teldo1} T. A. S. Pereira, J. S. de Sousa, G. A. Farias, J. A. K. Freire, M. H. Degani, and V. N. Freire, Appl. Phys. Lett. \textbf{87}, 171904 (2005).
\bibitem{teldo2}T. A. S. Pereira, J. S. de Sousa, J. A. K. Freire, and G. A. Farias, J. Appl.Phys. \textbf{108}, 054311 (2010). 
\bibitem{teldo3}A. A. Sousa, T. A. S. Pereira, A. Chaves, J. S. de Sousa, and G. A. Farias, Appl. Phys. Lett. \textbf{100}, 211601 (2012).
\bibitem{stern}F. Stern, Phys. Rev. B \textbf{17}, 5009 (1978); ibid., Solid State Commun. \textbf{25}, 163 (1977).
\bibitem{Degani0} M. H. Degani, Appl. Phys. Lett. \textbf{59}, 57 (1991).
\bibitem{degani1} M. H. Degani and M. Z. Maialle Journal of Comput. and Theoret. Nanosc. \textbf{7}, 454, (2010).
\bibitem{degani2} M. Z. Maialle and M. H. Degani, Phys. Rev. B \textbf{83}, 155308, (2011).

\bibitem{vurgaftman}I. Vurgaftman, J. R. Meyer, and L. R. Ram-Mohan, J. Appl. Phys. \textbf{89}, 5815 (2001).
\bibitem{kosloff} R. Kosloff, J. Phys. Chem. \textbf{92}, 2087 (1988).
\bibitem{degani} M. H. Degani, Phys. Rev. B \textbf{66}, 233306 (2002).
\bibitem{leburton}M. H. Degani and J. P. Leburton, Phys. Rev. B \textbf{44}, 10901 (1991).
\bibitem{Chaves} A. Chaves, G. A. Farias, F. M. Peeters, and B. Szafran, Phys. Rev. B \textbf{80}, 125331 (2009).
\bibitem{kumagai}M. Kumagai and T. Takagahara, Phys. Rev. B \textbf{40}, 12359 (1989).

\bibitem{cook}T. E. Cook, Jr., C. C. Fulton, W. J. Mecouch, R. F. Davis, G. Lucovsky, and R. J. Nemanich, J. Appl. Phys. \textbf{94}, 7155 (2003).
\bibitem{korotkov}R. Y. Korotkov, J. M. Gregie, and B. W. Wessels, Appl. Phys. Lett. \textbf{80}, 1731 (2002).
\bibitem{wu1}C. I. Wu and A. Kahn, J. Appl. Phys. \textbf{86}, 3209 (1999).
\end{thebibliography}
\end{document}